\documentstyle[preprint,aps]{revtex}
\begin{document}
\draft
\preprint{\begin{tabular}{l}
YUMS 93-14\\SNUTP 93-43\\KEK-TH-366\\(July 1993)\end{tabular}}
\vskip 2.5cm
\title{\bf Anomalous Gauge-Boson Couplings\\
in High Energy $ep$ Collisions}
\author{
C. S. Kim$^{1,}$\footnote{kim@cskim.yonsei.ac.kr},
Jungil Lee$^{2,}$\footnote{jungil@phyy.snu.ac.kr},
and H. S. Song$^{2,}$\footnote{hssong@phyy.snu.ac.kr},}
\address{
$^{1}$ Department of Physics, Yonsei University, Seoul 120-749, Korea\\
$^{2}$ Center for Theoretical Physics and Department of Physics,\\
Seoul National University, Seoul 151-742, Korea}
\vskip 2.5cm
\maketitle
\begin{abstract}
We investigate the sensitivity of total cross sections of
$e+p \rightarrow W,Z$ to CP-conserving non-standard $WW\gamma$
couplings. We include all the important production mechanisms and
study the dependence of the total $W$ cross sections on the
anomalous $WW\gamma$ couplings, $\kappa$ and $\lambda$.
We argue that the ratio of $W$ and $Z$ production cross sections
is particularly well suited, being relatively insensitive to
uncertainties in the theoretical and experimental parameters.
\end{abstract}
\renewcommand{\thepage}{}
\newpage
\renewcommand{\thepage}{\arabic{page}}\setcounter{page}{1}
\begin{center}{\bf 1.~Introduction}\end{center}

Despite impressive experimental confirmation of the correctness
of the Standard Model~(SM), the most direct consequence of the
$SU(2) \times U(1)$ gauge symmetry, the nonabelian self-couplings
of $W,Z$, and photon remains poorly measured to date.
Futhermore, gauge boson coupling strengths are strongly constrained
by gauge invariance, and are sensitive to deviations from the SM. Hence,
experimental bounds on these couplings might shed light on new physics
beyond the SM.

In order to parametrize non-standard effects, it is important to know
what sort of additional couplings
can arise once the restrictions due to gauge
invariance are lifted. As has been previously shown\cite{Hagiwara},
there can be $14$ or more non-standard couplings in the most general
case. To keep the analysis manageable, we restrict ourselves to C,~P
and $U(1)_{\mbox{em}}$ conserving couplings. This restriction leads to just two
anomalous form-factors, traditionally denoted by $\lambda$ and $\kappa$
in the $WW\gamma$ sector of the SM, which can be related to the anomalous
electric quadrupole and the anomalous magnetic dipole moment
of the $W$\cite{Aronson}. In the SM at tree level, $\lambda=0$ and
$\kappa=1$. At present the best experimental limits,
$-3.6<\lambda<3.5$ and $-3.5<\kappa<5.9$, are from a recent analysis
of the $W\gamma$ production at s$p\bar{p}$s
by UA(2) collaboration~[3]. While these bounds are compatible with the SM,
they are still too weak to really be considered as a precision test of the SM.
Futhermore, in the absence of beam polarization, it is unlikely that
there will be a significant improvement from the study of $W$ pair production
at LEP-II\cite{Zeppenfeld}.

The photoproduction of a single $W$ boson at $ep$ colliders will
provide a very precise test of the structure of the Standard Model
\hspace{0.1cm}$WW\gamma$ vertex. The situation there is much cleaner, for
example, than
in $pp$ or $p\bar{p}$ colliders, where a $W$ and a photon have to be
identified in the final state\cite{Alitti}.
Theoretical studies of the $WW\gamma$ vertex at $ep$ colliders
have been performed\cite{BauerZ,Kim} to investigate the possibility
of measuring the anomalous magnetic moment of the $W$.
The measurement of $\kappa$ at $ep$ colliders using the shape of the
$p_{_T}$ distribution of $W$ production at large $p_{_T}$ has been
previously investigated in\cite{BauerZ}. However, this method
suffers from the disadvantage of being sensitive to uncalculated
higher-order QCD corrections, uncertainties in the parton distribution
of the photon, experimental systematic uncertainties, etc~\cite{Bawa}. We have
previously found\cite{Kim} that a measurement of the anomalous coupling
in the $WW\gamma$ vertex at $ep$ colliders can best be achieved by
considering the ratio of the $W$ and $Z$ production cross sections.
The advantage of using a cross section ratio is that uncertainties
from the luminosity, structure functions, higher-order corrections,
QCD scale, etc. tend to cancel\cite{Kim}.

In the present study we investigate the possibility of measuring
both $\kappa$ and $\lambda$ at the same time by considering
the total cross sections of massive gauge bosons $W$ and $Z$
at $ep$ colliders. We include both the lowest-order resolved
processes and the dominant direct photoproduction processes. Care
must be taken to avoid double counting those phase-space regions
of the direct processes which are already included in the resolved
processes. In section~2 we first derive the
matrix-element-squared averaged over initial state
polarizations for the process
$\gamma+q \rightarrow q^{(\prime)}+V, V=W,Z,$
which is the main production mechanism of $W$ and $Z$ at $ep$ colliders.
Also we discuss the structure of the cross sections
by including all the possible processes.
In section~3 we present our numerical results for
the total cross sections as well as the ratio of $W$ and $Z$ cross sections.
It is interesting to note that the direct process
$\gamma+q \rightarrow q^{(\prime)}+W$
receives contributions from the triple-boson $WW\gamma$ vertex.
This then gives the possibility of studying the vertex to check the
Standard Model couplings.
Section~4 contains our conclusions.
\begin{center}{\bf 2.~Theoretical Details}\end{center}

If we restrict ourselves to C and P even couplings with electromagnetic
gauge invariance, the most general $WW\gamma$ vertex can be parametrized
in terms of an effective Lagrangian\cite{Hagiwara}
\begin{equation}
{\cal L}^{WW\gamma}_{eff}
=-ie
\left[
(W^\dagger_{\mu\nu}W^\mu-W^{\dagger\mu}W_{\mu\nu})A^\nu
+\kappa W^\dagger_{\mu}W_{\nu}F^{\mu\nu}
+\frac{\lambda}{m^2_W}
W^\dagger_{\rho\mu}{W^\mu}_{\nu}F^{\nu\rho}
\right],
\end{equation}
where $W^\mu$ and $A^\mu$ stand for the $W^-$ and the photon field,
respectively. The parameters $\kappa$ and $\lambda$ are
related~[2] to the magnetic dipole
moment~($\mu_{_W}$) and electric quadrupole moment~($Q_{_W}$) of the
$W^+$ by
\begin{equation}
\mu_{_W}=\frac{e}{2m_{_W}}(1+\kappa+\lambda)
\hspace{.2in}\mbox{and}\hspace{.2in}
Q_{_W}=-\frac{e}{m^2_{_W}}(\kappa-\lambda).
\end{equation}
At the tree level of the Standard Model,
the non-abelian gauge structure only
allows for $\kappa=1$ and $\lambda=0$.

We begin with a discussion of the matrix elements for the process
$\gamma+q \rightarrow q^{(\prime)}+V, V=W, Z$
which is the dominant mechanism for $W, Z$ production at high energy
$ep$ colliders. The relevant helicity amplitude may be obtained directly
from Ref.~\cite{BauerZ}. After squaring, and summing over the helicities,
and simplifying the resulting expression, we obtain the hard scattering
cross sections
\renewcommand{\theequation}{3\alph{equation}}\setcounter{equation}{0}
\begin{equation}
\left(\frac{d\hat{\sigma}} {d\hat{t}}\right)^{D}
(\gamma+q \rightarrow q^{(\prime)}+V)
=
\frac{1}{16\pi\hat{s}^{2}}
\Sigma |V|^{2},
\end{equation}
with
\begin{eqnarray}
\Sigma|Z|^{2}
&=&-(g_{_Z}^{2}e^{2}e_{q}^{2}g_{q}^{2})
T_{0}(\hat{u},\hat{t},\hat{s},m_{_Z}^{2})/2,\nonumber\\
\Sigma|W|^{2}
&=&
-(g^{2}e^{2}|V_{qq^{\prime}}|^{2})T(\hat{u},\hat{t},\hat{s},m_{_W}^{2}
,Q,\kappa,\lambda)/2,\nonumber\\
\Sigma|W|^{2}_{SM}
&=&-(g^{2}e^{2}|V_{qq^\prime}|^{2})
T(\hat{u},\hat{t},\hat{s},m_{_W}^{2},Q,1,0)/2\nonumber\\
&=&-(g^{2}e^{2}|V_{qq^\prime}|^{2})\left(Q
-\frac{\hat{s} }{ \hat{s}+\hat{t}}\right)^{2}
T_{0}(\hat{u},\hat{t},\hat{s},m_{_W}^{2})/2,\\
 \mbox{and}\hspace{0.4cm}Q=|e_{q}|&,&
\hspace{0.4cm}g^2_{q}=\frac{1}{2}(1-4Qx_{_W}+8Q^{2}x^2_{_W}),
\hspace{0.4cm}x_{_W}=0.23,\nonumber
\end{eqnarray}\label{eq:dsdt}
where the subscript SM denotes the Standard Model parametrization
with $\kappa=1$, $\lambda=0$, and where
\begin{eqnarray}
&T_{0}&(\hat{s},\hat{t},\hat{u},{m_{_V}}^{2})=
\left(
\frac{{\hat{t}}^{2}+{\hat{u}}^{2}+2\hat{s}m_{_V}^{2}
}{\hat{t}\hat{u}}
\right),\nonumber\\
&T&(\hat{s},\hat{t},\hat{u},m_{_W}^{2},Q,\kappa,\lambda)
=
(Q-1)^{2}\frac{\hat{u} }{ \hat{t}}
+Q^{2}\frac{\hat{t} }{ \hat{u}}
+2Q(Q-1)m_{_W}^{2}\frac{\hat{s} }{ \hat{u}\hat{t}}
\nonumber\\
&&
-\left(
(Q-1)\frac{1}{ \hat{t}}-Q\frac{1}{\hat{u}}
\right)
(2\hat{s}m_{_W}^{2}-(1+\kappa)\hat{u}\hat{t})
\frac{1}{m_{_W}^{2}-\hat{s}}
+\frac{\hat{s}}{2m_{_W}^{2}}\\
&&-\left(
2\hat{u}(\hat{u}+\hat{s})\frac{1}{m_{_W}^{2}}
+(1+\kappa)
\left[
\hat{s}-(\hat{u}+\hat{s})^{2}\frac{1}{m_{_W}^{2}}
\right]
\right)
\frac{1}{2 (m_{_W}^{2}-\hat{s})}
\nonumber\\
\end{eqnarray}
\newpage
\begin{eqnarray}
&&
+\left(
8{\hat{u}}^{2}-16\hat{s}m_{_W}^{2}-4(1+\kappa){\hat{u}}^{2}
\left[1+\frac{\hat{s}}{m_{_W}^{2}}\right]\right.
\nonumber\\
&&\hspace{.7cm}
\left.+(1+\kappa)^{2}
\left[
4\hat{u}\hat{t}+({\hat{u}}^{2}+{\hat{t}}^{2})
\frac{\hat{s}}{m_{_W}^{2}}
\right]
\right)
\frac{1}{8(m_{_W}^{2}-\hat{s})^{2}}
\nonumber\\
&&
-{\lambda}^{2}
\frac{\hat{s}\hat{t}\hat{u}}{2m_{_W}^{4}(m_{_W}^{2}-\hat{s})}
+\lambda(2\kappa+\lambda-2)\frac{\hat{s}}{8m_{_W}^{2}}
\left[
1+\frac{2\hat{t}\hat{u}}{(m_{_W}^{2}-\hat{s})^{2}}
\right].\nonumber
\end{eqnarray}
By setting the quark charge $Q=|e_{q}|=1$, we can obtain the matrix
elements for the processes, $\gamma+e \rightarrow \nu+W$ and
$\gamma+e \rightarrow e+Z$. With the definitions of
$Y=\hat{s}/4m_{_W}^{2}, X=(Y-1/4)(1+\cos\theta)/2$ and
$\chi=1-\kappa$, the differential cross section with
respect to $\hat{\theta}$, the angle between the outgoing $W$ and the
incoming photon is
\renewcommand{\theequation}{4\alph{equation}}\setcounter{equation}{0}
\begin{equation}
\frac{\hat{d\sigma}}{d\cos\hat{\theta}}
(\gamma+q \rightarrow q^{\prime}+W)=
\frac{\pi\alpha^{2}(Y-1/4)}
{128m_{_W}^{2}Y^{2}(Y-X)^{2}\sin^{2}\theta_{W}}
F(Q=|e_{q}|)
,
\end{equation}
where
\begin{eqnarray}
F(Q)&=&
X\left[8Y-4+(8X^{2}+4X+1)/Y\right]\nonumber\\
&&-8\chi{X}(Y+X)-32\lambda(\lambda-\chi)YX(Y-X)
+64\lambda^{2}YX(Y-X)^{2}\\
&&+(\lambda-\chi)^{2}
\left[(Y^{2}+X^{2})(4Y-4X-1)+4XY\right]\nonumber\\
&&+8\xi_{_Q}
\left[-\chi(Y+X)+(\xi_{_Q}+2X)f\right],\nonumber
\end{eqnarray}
with\\
{\hskip 2cm}$\xi_{_Q}=(Y-X)(1-Q)
\hspace{0.3cm} \mbox{and}\hspace{0.3cm}
f=\left[(Y-1/4)^{2}+(X+1/4)^{2}\right]/\left(XY\right).$\\
The function $F(Q=1)$ repersents the matrix element for
$Q=|e_{q}|=1$, i.e. $\gamma+e \rightarrow \nu+W$.
In the Standard Model~ i.e.~ $\lambda=\chi=0$, $F(Q=1)$ vanishes
when the outgoing $W$ and the incoming photon are antiparallel~(~$X=0$~).
And that is the famous radiation zero~\cite{Mikaelian}.
It is interesting to note that the radiation zero is not a unique
feature of the Standard Model. The radiation zero will be
present~\cite{Abraham} whenever
\renewcommand{\theequation}{\arabic{equation}}\setcounter{equation}{4}
\begin{equation}
\lambda+\kappa=1\hspace{0.5cm}(~\mbox{or}~\lambda=\chi~)
\hspace{0.5cm}~\mbox{and}~\hspace{0.5cm}X=0
\end{equation}
for the process $\gamma+e \rightarrow \nu+W$.

Next we focus on the total production of $W$ and $Z$ in $ep$ collisions.
In the short term these processes will be studied at
HERA($E_{e}=30~$GeV$, E_{p}=820~$GeV$, {\cal L}=200~$pb$^{-1}~$yr$^{-1}$),
while in the long term availability of
LEP~$\times$~LHC(
$E_{e}=50~$GeV$, E_{p}=8000~$GeV$, {\cal L}=1000~$pb$^{-1}~$yr$^{-1}$)
collider will give collision energies in excess of 1~TeV.
We first calculate the total cross sections for the five different processes
which contribute to single $W$ and $Z$ production at $ep$ colliders.
{}From the sum of these contributions we then calculate the ratio
$\sigma_{total}(W)/\sigma_{total}(Z)$ as a function of the anomalous
$WW\gamma$ coupling parameters $\kappa$ and $\lambda$.
The five processes are
\renewcommand{\theequation}{6\alph{equation}}\setcounter{equation}{0}
\begin{eqnarray}
e^{-}+p &\rightarrow& e^{-}+W^{\pm}+X,\\
&\rightarrow& \nu+W^{-}+X,\\
&\rightarrow& e^{-}+Z+X ~(Z\mbox{ from hadronic vertex}),\\
\label{eq:Zh}
&\rightarrow& e^{-}+Z+X ~(Z\mbox{ from leptonic vertex}),\\
\label{eq:Zl}
&\rightarrow& \nu+Z+X.
\end{eqnarray}
The largest contributions for $W$ and $Z$ productions come from
the processes (6a) and (6c) which are dominated by
the real photon exchange Feynman diagrams with a photon emitted from
the incoming electron,
$e^{-}+p \rightarrow \gamma_{/e}+p \rightarrow V+X$,
and have been partly studied as a function of $\kappa$
in Ref.~\cite{Kim}.
The dominant subprocesses for
$\gamma+p \rightarrow V+X$ would appear to be the lowest order
$\bar{q}^{(\prime)}_{/\gamma}+q \rightarrow V$, where $q_{/\gamma}$
is a quark inside the photon. However this may not be strictly true,
even at very high energies, since quarks inside the photon $q_{/\gamma}$
exist mainly through the evolution $\gamma \rightarrow q\bar{q}$.
Hence the direct process
$\gamma+q \rightarrow q^{(\prime)}+V$
could be competitive with the lowest order contribution
$\bar{q}^{(\prime)}+q \rightarrow V$.
This raises the subtle question of double counting \cite{Kim,Blumlein}.
Certain kinematic regions of the direct processes contribute to the
evolution of $q_{/\gamma}$ which is already included in the lowest
order process. Both double counting and the mass singularities are
removed \cite{Olness} if we subtract the contribution of
$\gamma+q \rightarrow q^{(\prime)}+V$
in which the $\hat{t}$-channel-exchanged quark is on-shell
and collinear with the parent photon. Thus the singularity subtracted
lowest order contribution
from the subprocesses
$\bar{q}^{(\prime)}_{/\gamma}+q\rightarrow V$ is
\renewcommand{\theequation}{7\alph{equation}}\setcounter{equation}{0}
\begin{eqnarray}
\sigma^{L}(&e^{-}&+~p \rightarrow \gamma_{/e}+p \rightarrow V+X)
=\frac{{\cal C}^{L}_{_V}}{s}\int_{m^{2}_{_V}/s}^{1}
\frac{dx_{1}}{x_{1}}\nonumber\\
&&\times\left[\sum_{qq^{\prime}}
(f_{q_{/e}}-\tilde{f}_{q_{/e}})(x_1,m^{2}_{_V})
f_{q^\prime_{/p}}(\frac{m^{2}_{_V}}{x_{1}s},m^{2}_{_V})
+(q \leftrightarrow q^\prime)\right],
\end{eqnarray}
where
\begin{equation}
{\cal C}^{L}_{_W}=\frac{2{\pi}G_{_F}m^{2}_{_W}}{3\sqrt{2}}|V_{qq^\prime}|^{2},
\hspace{.5in}
{\cal C}^{L}_{_Z}=\frac{2{\pi}G_{_F}m^{2}_{_Z}}{3\sqrt{2}}g_{q}^{2}.
\end{equation}
The electron structure functions $f_{q/e}$
are obtained as usual
\renewcommand{\theequation}{\arabic{equation}}\setcounter{equation}{7}
\begin{eqnarray}
f_{q_{/e}}(x,Q^{2})&=&
\int_{x}^{1}\frac{dy}{y}
f_{q_{/\gamma}}(\frac{x}{y},Q^{2})f_{\gamma/e}(y),
\end{eqnarray}
where $f_{\gamma_{/e}}$ is the appropriate
Weiz\"{a}cker-Williams approximation \cite{Weizacker} of (quasi-real)
photon radiation, and ${f}_{q_{/\gamma}}$ is the usual photon structure
function. The part of photon structure function,
$\tilde{f}_{q_{/\gamma}}$, which results from photon splitting at large
$x$ (with large momentum transfer), has the leading order form as
\begin{eqnarray}
\tilde{f}^{(0)}_{q_{/\gamma}}(x,Q^{2})&=&
\frac{3\alpha{e}^{2}_{q}}{2\pi}
(1-2x+2x^{2})\log\left(\frac{Q^{2}}{\Lambda^{2}}\right),\nonumber\\
\mbox{and as before}\hspace{0.4cm}
\tilde{f}_{q_{/e}}(x,Q^{2})&=&
\int_{x}^{1}\frac{dy}{y}
\tilde{f}^{(0)}_{q_{/\gamma}}(\frac{x}{y},Q^{2})f_{\gamma/e}(y)\hspace{0.1cm}.
\end{eqnarray}

To obtain the total contribution from the direct subprocess,
$\gamma+q \rightarrow q^{(\prime)}+V$,
we must integrate Eq.~(3), regularizing the $\hat{t}$-pole
of the collinear singularity by cutting at the scale $\Lambda^{2}$
which determines the running of the photon structure functions
$f_{i_{/\gamma}}$. This corresponds to the subtraction used to redefine
the photon structure functions in Eq.~(7a). Then the
 hard scattering cross
sections from the direct subprocesses are
\renewcommand{\theequation}{10\alph{equation}}\setcounter{equation}{0}
\begin{equation}
\hat{\sigma}(\gamma+q \rightarrow q^{(\prime)}+V)
=\frac{{\cal C}^{D}_{_V}}{\hat{s}}\eta_{_V},
\end{equation}
where
\begin{eqnarray}
&\eta&_{_Z}(\hat{s},m^{2}_{_Z},\Lambda^{2})
=(1-2z+2z^{2})\log\left(\frac{\hat{s}-m^{2}_{_Z}}{\Lambda^{2}}\right)
+\frac{1}{2}(1+2z-3z^{2}),\nonumber\\
&\eta&_{W}(\hat{s},m^{2}_{W},\Lambda^{2},Q=|e_{q}|,\kappa,\lambda)
=(Q-1)^{2}(1-2z+2z^{2})
\log
\left(
\frac{\hat{s}-m^{2}_{W}}{\Lambda^{2}}
\right)
\\
&&\hspace{.15in}
-\left[(1-2z+2z^{2})-2Q(1+\kappa+2z^{2})
+\frac{(1-\kappa)^{2}}{4z}-\frac{(1+\kappa)^{2}}{4}
\right]\log{z}
\nonumber\\
&&\hspace{0.15in}
+\left[\left(2\kappa+\frac{(1-\kappa)^{2}}{16}\right)\frac{1}{z}
+\left(\frac{1}{2}+\frac{3(1+Q^{2})}{2}\right)z
+(1+\kappa)Q-\frac{(1-\kappa)^{2}}{16}+\frac{Q^{2}}{2}
\right](1-z)\nonumber\\
&&\hspace{.15in}
-\frac{\lambda^{2}}{4z^{2}}(z^{2}-2z\log{z}-1)
+\frac{\lambda}{16z}(2\kappa+\lambda-2)
\left[(z-1)(z-9)+4(z+1)\log{z}\right],\nonumber\\
\mbox{with}&&\nonumber\\
&&{\cal C}^{D}_{_W}=
\frac{\alpha{G}_{_F}m^{2}_{W}}{\sqrt{2}}|V_{qq^\prime}|^{2},
\hspace{.2in}
{\cal C}^{D}_{_Z}=\frac{\alpha{G}_{_F}m^{2}_{_Z}}{\sqrt{2}}g_{q}^{2}e_{q}^{2}
\hspace{.2in}\mbox{and}\hspace{.2in}z=\frac{m^{2}_{_V}}{\hat{s}}.
\end{eqnarray}
The first terms in the $\eta_{_{V=W,Z}}$ represent the collinear singularity
from the $\hat{t}$-pole exchange, which is related to the photon
structure-function of Eq.~(9). This is the singularity that has already
been subtracted in Eq.~(7), and so we can now add the two contributions
, Eqs.~(7) and (11), without double counting.
The total contribution from the direct subprocess
$\gamma+q\rightarrow q^{(\prime)}+V$
is
\renewcommand{\theequation}{\arabic{equation}}\setcounter{equation}{10}
\begin{eqnarray}
\sigma^{D}(&e^{-}&+~p \rightarrow \gamma_{/e}+p \rightarrow V+X)
=\frac{{\cal C}^{D}_{_V}}{s}
\int_{m^{2}_{_V}/s}^{1}\frac{dx_{1}}{x_{1}}
\int_{m^{2}_{_V}/x_{1}s}^{1}\frac{dx_{2}}{x_{2}}
\nonumber\\
&&\times
\left[\sum_{q}
f_{\gamma_{/e}}(x_1,Q^{2})f_{q_{/p}}(x_{2},Q^{2})
\right]
\eta_{_V}(\hat{s}=x_{1}x_{2}s).
\end{eqnarray}

The processes (6b) and (6d) are dominated by configurations
where a~(quasi-real) photon is emitted~(either elastically or
quasi-elastically) from the incoming proton and subsequently scatters
off the incoming electron, i.e.
$e^{-}+p \rightarrow e^{-}+\gamma_{/p} \rightarrow e^{-}~(~$or$~\nu)+V.$
For the elastic photon, the cross section can be computed using the
electrical and magnetic form factors of the proton.
For the quasi-elastic scattering photon, the experimental information
\cite{Stein} on electromagnetic structure functions $W_{1}$ and
$W_{2}$ can be used, following Ref.~\cite{BauerV}. The hard scattering
cross section
is given by
\begin{equation}
\hat{\sigma}(e^{-}+\gamma_{/p} \rightarrow e^{-}~(~\mbox{or}~\nu)+V)
=\frac{{\cal C}^{D}_{_V}}{\hat{s}}
\eta_{_V}(Q=|e_{q}|=1).
\end{equation}
For process (6e), which is a pure charged current process, we simply
use the results of Bauer {\it et. al.} \cite{BauerV} to add to the
contributions from (6c) and (6d). The contribution from this process
to the total $Z$ production cross section is almost negligible even at
LEP~$\times$~LHC $ep$ collider energies, as
can be seen in Table 2.

\begin{center}{\bf 3.~Numerical Results and Discussions}\end{center}

In Table 1 we show the total $W^{\pm}$ production cross section at HERA
and LEP~$\times$~LHC $ep$ colliders for a range of values of
 the anomalous $WW\gamma$
coupling parameters $\kappa$ and $\lambda$. The error range represents
the variation in the cross section by varying the theoretical input
parameters as follows~:~$m^{2}_{V}/10\leq{Q}^{2}\leq{m}^{2}_{V}$, photon
structure functions $f_{q_{/\gamma}}$ from DG~\cite{Drees} and
DO$+$VMD~\cite{Duke}, and proton structure functions $f_{q_{/p}}$
from EHLQ1~\cite{Eichten} and HMRS(B)~\cite{Harriman}.
It is important to note that
once photoproduction experiments at
HERA determine $f_{q_{/p}}$ and $f_{q_{/\gamma}}$ more precisely,
we will be able to predict the total cross sections for each process
with much greater accuracy.
The subtraction terms $\tilde{f}_{q_{/\gamma}}$ of Eq.~(7) have been
calculated using the leading order photon splitting function
as in Eq.~(9).
We show in Table 2 the cross sections
for the various $Z$ production channels
at the HERA and LEP~$\times$~LHC $ep$ colliders. The errors represent
the variation in cross sections obtained by varying the input
parameters, as in Table 1. With the anticipated luminocities of
${\cal L}=200~$pb$^{-1}$yr$^{-1}$~(HERA) and
${\cal L}=1000~$pb$^{-1}$yr$^{-1}$~(LEP~$\times$~LHC),
the total $Z$ production cross section corresponds to
84~events/yr~(HERA) and 5400~events/yr~(LEP~$\times$~LHC).
After including a $6.7~\%$ leptonic branching
ratio~(i.e.~$Z \rightarrow e^{+}e^{-},\mu^{+}\mu^{-}$),
the event numbers become about 6~events/yr(HERA) and
360~events/yr(LEP~$\times$~LHC).

In Table 3 we show the ratio
$\sigma(W^{+})/\sigma(Z)$, $\sigma(W^{-})/\sigma(Z)$ and
$\sigma(W^{+}+W^{-})/\sigma(Z)$ for the various values of $\kappa$
and $\lambda$. The input parameters have beeen varied as in Table~1.
Note also that we have not included the uncertainties due to higher
order perturbative QCD corrections. While these are expected to have
non-negligible effect on the absolute $W$ and $Z$
cross sections - as in $pp$ and $p\bar{p}$ collisions - they are
expected to largely cancel in the $W/Z$ cross section ratio, since
to a first approximation the gluons are blind to the quark flavor.
In Fig.~1, the ratio of $\sigma(W^{\pm})/\sigma(Z)$ and
$\sigma(W^{-})/\sigma(Z)$ is
shown as a function of $\kappa$ and $\lambda$.
Rather than vary both parameters simultaneously, we first set
$\kappa$ to its Standard Model value and then vary $\lambda$ and
vice versa. Assuming $\kappa=1$ as in Table~1 and Fig~1(a),
 the $W$ production
cross section and the ratio  $\sigma(W^{\pm})/\sigma(Z)$ may be used
to determine $\lambda$ up to a sign, since when $\kappa=1$,
the dependence on $\lambda$ is quadratic, i.e.
$\sigma(W^{\pm})_{\kappa=1}\propto \lambda^{2}$,
as can be easily seen in Eqs.~(3) and (4). Hence the sign ambiguity
can be resolved only by studying a separate process, such as $W$
pair production in LEP-II ~$e^{+}e^{-}$~ collider.
Notice also from Fig~1(b) that
$\sigma(W^{\pm})_{\lambda=0}\propto a\kappa^{2}+b\kappa$,
and the $W$ cross section has a minimum at $\kappa\approx-0.5$.
This means that there is another value $\kappa\approx-2$
which gives the same cross section as the Standard Model value $\kappa=1$.
In Fig.~2, we show three dimensional bar charts of ratio
$\sigma(W^{\pm})/\sigma(Z)$ at HERA, and
$\sigma(W^{-})/\sigma(Z)$ at LEP~$\times$~LHC.
The theoretical input parameters have been fixed : $Q^{2}=m^{2}_{V}$,
$f_{q_{/\gamma}}$ from DG~\cite{Drees}, and
$f_{q_{/p}}$ from EHLQ1~\cite{Eichten}.

To obtain an experimentally measurable ratio
$\sigma(ep \rightarrow W^{\pm} \rightarrow l\nu)/
\sigma(ep \rightarrow Z \rightarrow l^{+}l^{-})$
we must multiply the cross section ratio
$\sigma(W)/\sigma(Z)$ by the leptonic branching ratio factor
\begin{equation}
R_{BR}(m_{t}>m_{W}-m_{b},N_\nu=3)\equiv
\frac{BR(W^{\pm} \rightarrow l\nu)}{BR(Z^{\pm} \rightarrow l^{+}l^{-})}
=3.23.
\end{equation}
After 5 years of running, HERA will produce about 30
$e+p \rightarrow Z+X \rightarrow l^{+}+l^{-}+X$ events,
and this will enable us to determine $\kappa$ and $\lambda$
with a precision of order
\begin{eqnarray}
\Delta\kappa&\approx&{\pm}0.3\hspace{.1in}
\mbox{for}\hspace{.1in}\lambda=0,
\nonumber\\
\Delta\lambda&\approx&{\pm}0.8\hspace{.1in}
\mbox{for}\hspace{.1in}\kappa=1,
\end{eqnarray}
which are comparable with the expected constraints from
the future LEP-II ~$e^{+}e^{-}$~ experiment.
At LEP~$\times$~LHC, one year's running will give
\begin{eqnarray}
\Delta\kappa&\approx&{\pm}0.2\hspace{.1in}
\mbox{for}\hspace{.1in}\lambda=0,
\nonumber\\
\Delta\lambda&\approx&{\pm}0.3\hspace{.1in}
\mbox{for}\hspace{.1in}\kappa=1.
\end{eqnarray}
\begin{center}{\bf 4.~Conclusion}\end{center}

We have shown how measurements of weak boson production at high energy
electron-proton colliders can provide important information on anomalous
$WW\gamma$ couplings. We have analyzed the production of massive gauge
bosons - $W$ and $Z$. We have included both direct and indirect processes,
involing the parton structure of the photon, taking careful account of the
double counting problem for the latter.
We have also argued that the ratio of $W$ and $Z$ production cross sections
is particularly suited to an experimental determination of the anomalous
$WW\gamma$ coupling parameters $\kappa$ and $\lambda$, being relatively
insensitive to uncertainties in the theoretical input parameters.
In fact, with more precise measurements of these parameters in the next few
years - in particular the photon structure functions - the errors in the
measured $\kappa$ and $\lambda$ values will ultimately be obtained by
the statistical error from the small number of $Z$ events.
In this respect, the higher energy LEP~$\times$~LHC collider offers a
significant improvement. Finally we note that our estimated precision on
$\kappa$ and $\lambda$ for both $ep$ colliders,
Eqs.~(14) and (15), is an order of magnitude greater than existing
measurements from $W\gamma$ production
at $p\bar{p}$ collider~\cite{Alitti}.

Attempts are at present under way by many authors to constrain the
parameter space of $\lambda$ and $\kappa$ by considering
various experimental results; production of $W+\gamma$ at $p\bar{p}$
collider~[19], process $\gamma e\rightarrow W\nu$
at future $e^+e^-$ and $e\gamma$ colliders~[9,20], and also from
present low energy data~[21].
And those approaches should be regarded as complementary
in the efforts to find new physics beyond the Standard Model.
\begin{center}{\bf Acknowledgements}\end{center}

CSK would like to thank K.~Hagiwara and the hospitality of KEK,
where the part of this work has been completed.
The work was supported in part by the Korea Science and Engineering
Foundation and in part by the Korean Ministry of Education.
The work of CSK was also supported in part by the Center
for Theoretical Physics at Seoul National University and in part by a
Yonsei University Faculty Research Grant.

\newpage
\renewcommand{\thepage}{T-\arabic{page}}\setcounter{page}{1}
\begin{center}
{\bf Table Captions}
\end{center}

Table 1. Total $W$-production cross sections (in pb) at HERA and at
LEP~$\times$~LHC, as a function of anomalous $WW\gamma$ coupling
parameters $\kappa$ and $\lambda$. The error range represents the
uncertainties in the cross sections by varying the theoretical input
parameters : $m^{2}_{V}/10 \leq Q^{2} \leq m^{2}_{V}$,
photon structure functions $f_{q_{/\gamma}}$~(DG[15] and DO+VMD[16]),
and proton structure functions $f_{q_{/p}}$~(EHLQ1[17] and HMRS(B)[18]).
\newline

Table 2. The cross sections~(in pb) for the various $Z$ production channels
at HERA and LEP~$\times$~LHC $ep$ colliders.
The errors represent the variation in cross sections obtained by
varying the theoretical input parameters, as in Table~1.
\newline

Table 3. Production cross section ratio of $W/Z$ as a function of $\kappa$ and
$\lambda$ at HERA and LEP~$\times$~LHC. We first set $\lambda$ to
its Standard Model values~($\lambda=0$) and then vary $\lambda$
and vice versa.

\newpage
\begin{center}
\begin{tabular}{|c|c|c|c|}
\multicolumn{4}{c}{\bf HERA $W$-production Cross-section~(in pb)} \\ \hline
&$ep\rightarrow{W^+}X$& $ep\rightarrow{W^-}X$& $ep\rightarrow{W^\pm}X$\\ \hline
\makebox[4cm]{$\lambda=0$, $\kappa=0.0$}
&\makebox[3.5cm]{0.43 $\pm$ 0.08}
&\makebox[3.5cm]{0.47 $\pm$ 0.08}
&\makebox[3.5cm]{0.86 $\pm$ 0.12}\\\hline
$\lambda=0$, $\kappa=0.5$&0.49 $\pm$ 0.08&0.51 $\pm$ 0.07&0.97 $\pm$ 0.12
\\\hline
$\lambda=0$, $\kappa=1.0$&0.59 $\pm$ 0.08&0.59 $\pm$ 0.07&1.15 $\pm$ 0.12
\\\hline
$\lambda=0$, $\kappa=1.5$&0.72 $\pm$ 0.08&0.70 $\pm$ 0.08&1.39 $\pm$ 0.12
\\\hline
$\lambda=0$, $\kappa=2.0$&0.88 $\pm$ 0.08&0.85 $\pm$ 0.09&1.69 $\pm$ 0.13
\\\hline
$\lambda=0.0$, $\kappa=1$&0.59 $\pm$ 0.08&0.59 $\pm$ 0.08&1.15 $\pm$ 0.12
\\\hline
$\lambda=0.5$, $\kappa=1$&0.60 $\pm$ 0.08&0.61 $\pm$ 0.08&1.17 $\pm$ 0.12
\\\hline
$\lambda=1.0$, $\kappa=1$&0.63 $\pm$ 0.08&0.64 $\pm$ 0.08&1.23 $\pm$ 0.12
\\\hline
$\lambda=1.5$, $\kappa=1$&0.68 $\pm$ 0.08&0.68 $\pm$ 0.08&1.32 $\pm$ 0.13
\\\hline
$\lambda=2.0$, $\kappa=1$&0.75 $\pm$ 0.08&0.74 $\pm$ 0.07&1.46 $\pm$ 0.13
\\\hline
\multicolumn{4}{c}{}\\
\multicolumn{4}{c}{\bf LEP$\times$LHC $W$-production Cross-section~(in pb)}
\\ \hline
&$ep\rightarrow{W^+}X$& $ep\rightarrow{W^-}X$& $ep\rightarrow{W^\pm}X$
\\ \hline
\makebox[4cm]{$\lambda=0$, $\kappa=0.0$}
&\makebox[3.5cm]{6.94 $\pm$ 2.09}
&\makebox[3.5cm]{8.05 $\pm$ 0.82}
&\makebox[3.5cm]{14.99 $\pm$ 2.91}
\\\hline
$\lambda=0$, $\kappa=0.5$&8.65 $\pm$ 1.65&9.49 $\pm$ 0.70&18.13 $\pm$ 2.36
\\\hline
$\lambda=0$, $\kappa=1.0$&11.24 $\pm$ 1.46&12.12 $\pm$ 0.86&23.36 $\pm$ 2.33
\\\hline
$\lambda=0$, $\kappa=1.5$&14.67 $\pm$ 1.47&16.14 $\pm$ 1.02&30.81 $\pm$ 2.49
\\\hline
$\lambda=0$, $\kappa=2.0$&19.06 $\pm$ 1.80&21.34 $\pm$ 1.37&40.40 $\pm$ 3.17
\\\hline
$\lambda=0.0$, $\kappa=1$&11.24 $\pm$ 1.46&12.13 $\pm$ 0.85&23.37 $\pm$ 2.31
\\\hline
$\lambda=0.5$, $\kappa=1$&12.87 $\pm$ 1.46&14.15 $\pm$ 0.95&27.02 $\pm$ 2.41
\\\hline
$\lambda=1.0$, $\kappa=1$&17.79 $\pm$ 1.49&20.69 $\pm$ 0.81&38.47 $\pm$ 2.29
\\\hline
$\lambda=1.5$, $\kappa=1$&25.97 $\pm$ 1.51&31.12 $\pm$ 1.09&57.08 $\pm$ 2.60
\\\hline
$\lambda=2.0$, $\kappa=1$&37.52 $\pm$ 1.65&45.89 $\pm$ 1.26&83.40 $\pm$ 2.91
\\\hline
\multicolumn{4}{c}{}\\
\multicolumn{4}{c}{Table 1.}\\
\end{tabular}
\end{center}
\newpage
\begin{center}
\begin{tabular}{|l|c|c|}
\multicolumn{3}{c}{\bf $Z$-production Cross-sections~(in pb)} \\ \hline
\multicolumn{1}{|c|}{Process}   &HERA           &LEP$\times$LHC \\ \hline
\hspace{2cm}\makebox[5.3cm][l]{$ep$ ${\rightarrow}$ $eZX$ (hadronic)}
&\makebox[3.6cm]{0.25 $\pm$ 0.05}
&\makebox[3.6cm]{3.61 $\pm$ 0.59}\\ \hline
\hspace{2cm}$ep$ ${\rightarrow}$ $eZX$ (leptonic)  &0.16&1.17\\ \hline
\hspace{2cm}$ep$ ${\rightarrow}$ ${\nu}ZX$        &0.004&0.61\\ \hline
\hspace{2cm}$ep$ ${\rightarrow}$ $ZX$
&0.42 $\pm$ 0.05  &5.39 $\pm$ 0.59\\ \hline
\multicolumn{3}{c}{}\\
\multicolumn{3}{c}{Table 2.}\\
\end{tabular}
\end{center}
\newpage
\begin{center}
\begin{tabular}{|c|c|c|c|}
\multicolumn{4}{c}{\bf HERA $W$-production Ratio} \\ \hline
&{$\sigma(W^+)/\sigma(Z)$}
&{$\sigma(W^-)/\sigma(Z)$}
& {$\sigma(W^\pm)/\sigma(Z)$}\\ \hline
\makebox[4cm]{$\lambda=0$, $\kappa=0.0$}
&\makebox[3.5cm]{1.12 $\pm$ 0.11}
&\makebox[3.5cm]{1.29 $\pm$ 0.16}
&\makebox[3.5cm]{2.32 $\pm$ 0.18}\\\hline
$\lambda=0$, $\kappa=0.5$&1.29 $\pm$ 0.10&1.41 $\pm$ 0.14&2.60 $\pm$ 0.16
\\\hline
$\lambda=0$, $\kappa=1.0$&1.55 $\pm$ 0.09&1.61 $\pm$ 0.15&3.11 $\pm$ 0.16
\\\hline
$\lambda=0$, $\kappa=1.5$&1.90 $\pm$ 0.07&1.91 $\pm$ 0.17&3.78 $\pm$ 0.15
\\\hline
$\lambda=0$, $\kappa=2.0$&2.34 $\pm$ 0.05&2.28 $\pm$ 0.21&4.61 $\pm$ 0.20
\\\hline
$\lambda=0.0$, $\kappa=1$&1.55 $\pm$ 0.09&1.62 $\pm$ 0.16&3.12 $\pm$ 0.16
\\\hline
$\lambda=0.5$, $\kappa=1$&1.58 $\pm$ 0.08&1.67 $\pm$ 0.18&3.19 $\pm$ 0.18
\\\hline
$\lambda=1.0$, $\kappa=1$&1.66 $\pm$ 0.08&1.73 $\pm$ 0.17&3.34 $\pm$ 0.17
\\\hline
$\lambda=1.5$, $\kappa=1$&1.79 $\pm$ 0.07&1.86 $\pm$ 0.18&3.63 $\pm$ 0.19
\\\hline
$\lambda=2.0$, $\kappa=1$&1.99 $\pm$ 0.06&2.00 $\pm$ 0.16&3.97 $\pm$ 0.14
\\\hline
\multicolumn{4}{c}{}\\
\multicolumn{4}{c}{\bf LEP$\times$LHC $W$-production Ratio} \\ \hline
&{$\sigma(W^+)/\sigma(Z)$}
&{$\sigma(W^-)/\sigma(Z)$}
&{$\sigma(W^\pm)/\sigma(Z)$}\\ \hline
\makebox[4cm]{$\lambda=0$, $\kappa=0.0$}
&\makebox[3.5cm]{1.38 $\pm$ 0.28}
&\makebox[3.5cm]{1.62 $\pm$ 0.06}
&\makebox[3.5cm]{3.00 $\pm$ 0.27}\\\hline
$\lambda=0$, $\kappa=0.5$&1.73 $\pm$ 0.15&1.90 $\pm$ 0.09&3.68 $\pm$ 0.11
\\\hline
$\lambda=0$, $\kappa=1.0$&2.27 $\pm$ 0.06&2.43 $\pm$ 0.12&4.71 $\pm$ 0.18
\\\hline
$\lambda=0$, $\kappa=1.5$&2.97 $\pm$ 0.12&3.27 $\pm$ 0.19&6.24 $\pm$ 0.31
\\\hline
$\lambda=0$, $\kappa=2.0$&3.89 $\pm$ 0.20&4.38 $\pm$ 0.27&8.27 $\pm$ 0.47
\\\hline
$\lambda=0.0$, $\kappa=1$&2.27 $\pm$ 0.06&2.43 $\pm$ 0.12&4.71 $\pm$ 0.18
\\\hline
$\lambda=0.5$, $\kappa=1$&2.61 $\pm$ 0.07&2.87 $\pm$ 0.15&5.48 $\pm$ 0.22
\\\hline
$\lambda=1.0$, $\kappa=1$&3.62 $\pm$ 0.13&4.22 $\pm$ 0.28&7.82 $\pm$ 0.37
\\\hline
$\lambda=1.5$, $\kappa=1$&5.28 $\pm$ 0.25&6.35 $\pm$ 0.45&11.63 $\pm$ 0.70
\\\hline
$\lambda=2.0$, $\kappa=1$&7.65 $\pm$ 0.47&9.37 $\pm$ 0.74&17.01 $\pm$ 1.20
\\\hline
\multicolumn{4}{c}{}\\
\multicolumn{4}{c}{Table 3.}\\
\end{tabular}
\end{center}
\newpage
\renewcommand{\thepage}{F-\arabic{page}}\setcounter{page}{1}
\begin{center}
{\bf Figure Captions}
\end{center}

Fig.~1. Total production cross section ratios
$\sigma(W^{-})/\sigma(Z)$ and $\sigma(W^{+}+W^{-})/\sigma(Z)$
as a function of
(a)~$\lambda~(\kappa=1)$, and
(b)~$\kappa~(\lambda=0)$ at the HERA and LEP~$\times~$LHC
$ep$ colliders.
The errors represent the variation in cross sections obtained by
varying the theoretical input parameters,
as in Table~1. The experimentally measured ratio~
$\sigma(ep \rightarrow W \rightarrow l\nu)/
\sigma(ep \rightarrow Z \rightarrow l^{+}l^{-})~(l=e,\mu)$
is obtained by multiplying by the leptonic branching ratio factor
$BR(W \rightarrow l\nu)/BR(Z \rightarrow l^{+}l^{-})=3.23$,
assuming $m_{t}>75$~GeV and three light neutrino species.
\newline

Fig.~2. Three dimensional bar chart of ratio
(a)~$\sigma(W^{+}+W^{-})/\sigma(Z)$ at the HERA, and
(b)~$\sigma(W^{-})/\sigma(Z)$ at LEP~$\times$~LHC.
Theoretical input parameters have been fixed :
$Q^{2}=m^{2}_{V}$, $f_{q_{/\gamma}}$ from DG~\cite{Drees}
and $f_{q_{/p}}$ from EHLQ1~\cite{Eichten}.
\end{document}